\documentclass[reprint,prl, showpacs, showkeys, amsmath, amssymb, floatfix, preprintnumbers, raggedcolumns]{revtex4-1}
\pdfoutput=1
\usepackage{graphicx}
\usepackage{enumerate}
\usepackage{bm}
\usepackage{slashed}
\usepackage{comment}
\usepackage{bbold}
\usepackage{braket}
\usepackage{makecell}
\usepackage{amssymb}
\usepackage{epsfig}
\usepackage{epstopdf}

\newcommand{\be}{\begin{equation}}
\newcommand{\ee}{\end{equation}}
\newcommand\beq{\begin{eqnarray}}
\newcommand\eeq{\end{eqnarray}} 
\newcommand\eqn[1]{\label{eq:#1}} 
\newcommand\eq[1]{Eq.~(\ref{eq:#1})}  
\newcommand\eqB[1]{(\ref{eq:#1})}

\newcommand{\CO}{{\cal O}}

\newcommand{\Tr}{{\rm Tr\,}}
\newcommand\half{{\textstyle{\frac{1}{2}}}}

\newcommand{\mybar}[1]%
        {\kern 0.6pt\overline{\kern -0.6pt#1\kern -0.6pt}\kern 0.6pt}
\makeatletter
\renewcommand*\env@matrix[1][\arraystretch]{%
  \edef\arraystretch{#1}%
  \hskip -\arraycolsep
  \let\@ifnextchar\new@ifnextchar
  \array{*\c@MaxMatrixCols c}}
\makeatother

\begin{document}

\preprint{INT-PUB-15-064}

\title{A  Nonperturbative Regulator for Chiral Gauge Theories}

\author{Dorota M. Grabowska}
 \email{grabow@uw.edu}

\author{David B. Kaplan}
\email{dbkaplan@uw.edu}
\affiliation{Institute for Nuclear Theory, Box 351550, Seattle, Washington 98195-1550, USA}
 
 \date{Received 1 December 2015; revised manuscript received 8 April 2016; published 24 May 2016}
 
 \begin{abstract}
 We propose a nonperturbative gauge invariant regulator for $d$-dimensional chiral gauge theories on the lattice.  The method involves simulating domain wall fermions in $d+1$ dimensions with quantum gauge fields that reside on one $d$-dimensional surface and are extended into the bulk via gradient flow.  The result is a theory of gauged fermions plus mirror fermions, where  the mirror fermions couple to the gauge fields via a form factor that becomes exponentially soft with the separation between domain walls. The resultant theory has a local $d$-dimensional interpretation  only if the chiral fermion representation is anomaly free. A physical realization of this construction would imply the existence of mirror fermions in the standard model that are invisible except for interactions induced by vacuum topology, and which could gravitate differently than conventional matter.
  
  \end{abstract}

\pacs{11.15.-q,11.15.Ha,71.10.Pm}
 \maketitle
\section{Introduction}
There is a fundamental tension between taming the ultraviolet behavior of a chiral gauge theory and maintaining gauge invariance. There is no perturbative regulator known to work to all orders, and constructing a nonperturbative lattice  regulator for chiral gauge theories in $d=2,4$ dimensions has been a daunting problem for decades \cite{Golterman:2000hr}.  This is of particular interest since the standard model is a chiral gauge theory.   The lack of a regulator may be a purely technical problem, but it might  indicate that something is missing from the standard model.  Naive lattice fermions are always Dirac in structure, leading to unwanted mirror fermions;  in the case of chiral gauge theories, decoupling these mirror fermions by means of a large mass would entail explicitly breaking the gauge symmetry.  Attempts to solve the problem generally fall into three classes: (i)  the  gauge symmetry is broken spontaneously \cite{Kaplan:1992sg} or explicitly  \cite{Golterman:2004qv} and mirror fermions are decoupled from the gauge fields, with a procedure  to recover gauge symmetry  in the continuum large-volume limit, (ii) the mirror fermions are given exotic  gauge-invariant strong interactions intended to induce a mass gap in that sector \cite{Swift:1984dg,Smit:1985nu,Eichten:1985ft,Poppitz:2007tu,Wen:2013ppa,you2015interacting}, or (iii) the mirror fermions are projected out of the theory \cite{Narayanan:1993sk,Luscher:1998du}.  In the first category, the spontaneous breaking of  gauge symmetry fails to yield a chiral fermion spectrum \cite{Golterman:1994at}, while  to date the explicit breaking approach has only been shown  to recover continuum gauge invariance  in perturbation theory.   The strategy of decoupling  mirror fermions via exotic interactions is a nontrivial dynamical question. In \cite{Kaplan:1992bt}  the gauge fields were given space-dependent couplings, but it is thought that the construction is unlikely  to have a continuum limit \cite{KorthalsAltes:1993dk}. Certain other cases of exotic interactions have been closely examined and do not appear to  possess the expected mass gap  \cite{Golterman:1991re,Golterman:1992yha,Chen:2012di}.    In the third category, where mirror fermions are projected out, one must show that the resulting fermion contribution to the gauge measure is analytic in the gauge fields and can be derived from a  local fermion action. Examples in this category include the ansatz  \cite{Narayanan:1993sk}  based on overlap fermions  \cite{Narayanan:1992wx} which is analytic, but not obviously derivable from a local fermion action, and  the work of Ref.~\cite{Luscher:1998du} which starts from the chirally projected overlap operator  \cite{Neuberger:1997fp} and provides an implicit construction of a local and analytic measure in the case of $U(1)$ gauge symmetry, but which has not been generalized to non-Abelian gauge symmetries.  Any nonperturbative solution is expected to agree with low order perturbation theory, including a path to failure for the case of anomalous fermion representations, and so this will be our criteria for a successful nonperturbative regulator.

The problem of how to realize global chiral symmetries on the lattice with the correct anomalies was resolved in Ref.~ \cite{Kaplan:1992bt}.  In this  construction  Dirac fermions in $2n+1$ dimensions are introduced; the theory possesses a mass gap in the bulk and massless chiral  modes localized on the ``domain walls", which are the $2n$-dimensional surfaces of the space. The number of such modes is a topological invariant of the bulk fermion dispersion relation \cite{Golterman:1992ub} and the theory is an example of what condensed matter physicists currently refer to as a topological insulator.  Chiral symmetry becomes exact  in the limit of infinite  extra dimension, in which case the effective $2n$-dimensional description of these modes is the overlap fermion \cite{Narayanan:1992wx,Neuberger:1997fp}.  The geometry of domain wall fermions naturally suggests  that by  localizing gauge interactions near one surface of the extra dimension one might obtain a continuum chiral gauge theory while mirror fermions on the distant surface  decouple;  this has been a starting point for many of the  attempts to construct chiral gauge theories  cited above.  There is some reason to be optimistic about a solution involving an extra dimension \cite{Kaplan:1995pe}, even though particular dynamical realizations have not been successful.  In this Letter we propose a gauge-invariant solution based on domain wall fermions that works on a new principle: gauge fields are extended into the extra dimension via ``gradient flow", as solutions to a gauge covariant parabolic differential equation.  The effect is that the mirror fermions are endowed with nonlocal but gauge-invariant couplings which allow them to  decouple in perturbation theory, leaving behind a local $d$-dimensional chiral gauge theory in the infrared when (and only when) the theory is gauge anomaly free.
 A striking feature of this proposal is that even in the anomaly-free case the mirror fermions are still sensitive to topological features of the gauge field and can therefore  have nonlocal and nonperturbative interactions with ordinary matter. For all we know, such interactions could be a necessary intrinsic feature of non-Abelian chiral gauge theories, implying exotic and yet to be discovered   phenomenology in the standard model that is not apparent in Feynman diagrams.

\section{Definition of the Chiral Measure}   

Euclidean Green  functions in a gauge theory can be expressed as path integral averages of functionals of gauge fields with respect to a measure $e^{-S(A)}\Delta(A)$ where
$S$ is the Maxwell or Yang-Mills action and $\Delta$ is the fermion contribution to the measure.   In a vectorlike gauge theory  $\Delta$ is just given by a product of one determinant of the Dirac operator for each fermion flavor; in a chiral gauge theory $|\Delta|^2$ must equal a product of Dirac determinants, but the  problem is how to define the phase of $\Delta$ such  that it is analytic in $A_\mu$ and follows from a local fermion action.  It is known that if the fermion representation is in an anomalous representation of the gauge symmetry, this phase will be gauge variant and the theory  ill-defined. Our proposal for $\Delta$ for a single Weyl fermion  in the continuum with dimension $d=2,4$ is 
\beq
\Delta(A)=\frac{\det\left[\slashed{D}^{(R)}_{d+1} - \Lambda \epsilon(s)\right]}{\det\left[\slashed{D}^{(R)}_{d+1} -\Lambda \right]}\ .
\eqn{smeasure}\eeq
In this expression $\slashed{D}^{(R)}_{d+1} $ is the $(d+1)$-dimensional Dirac operator in the gauge group representation $R$ for the fermion,  where the extra dimension denoted by coordinate $s\in [-L,L]$   is a circle with circumference $2L$, $\epsilon(s)=\text{sgn}(s)$,  and $\Lambda$ is a real mass scale whose sign is the fermion chirality.  The scale $|\Lambda|$ can be thought of as  the ultraviolet cutoff of the theory and will be equated with the inverse lattice spacing in a discretized version of the theory, with $|\Lambda | L\to\infty$.  

So far our expression for $\Delta$  would just seem to describe the determinants of a normal domain wall fermion, with zeromodes of chirality $\mp\text{sgn}(\Lambda)$ localized on the mass defects at $s=0$ and $s=\pm L$ respectively, with a Pauli-Villars field of constant mass $\Lambda$ which  cancels off unwanted effects of the  heavy bulk fermions.  What  differs in the present formulation is the gauge field in $\slashed{D}_{d+1}$.  In the usual application of domain wall fermions to lattice QCD the $d$-dimensional gauge fields are independent of the coordinate $s$. In contrast we specify here an $s$-dependent, $d$-dimensional gauge field  $\bar A_\mu(x,s)$  solving the gradient flow equation
\beq
\partial_s \bar A_\nu =\frac{\xi \epsilon(s)}{|\Lambda|} D_\mu \bar F_{\mu\nu}\ ,\quad \mu,\nu=1,\ldots,d\ ,
\eqn{flow}
\eeq
with boundary condition $ \bar A_\mu(x,0)= A_\mu(x)$, where $A_\mu(x)$ is the integration variable in the path integral. In the above equation   $D_\mu$ and $\bar F_{\mu\nu}$ are the covariant derivative and Yang-Mills (or Maxwell) field strength respectively constructed out of $\bar A_\mu(x,s)$, where the indices run to $d$, not $d+1$. The parameter $\xi$ is  dimensionless and can be set to unity for applications, but it is useful for our discussion to keep its value general, allowing us to interpolate between the conventional application of domain wall fermions with $\xi=0$ and  $s$-independent  gauge fields, and the case $\xi\gtrsim 1$ where the gauge flow is rapid.    Note that  \eq{flow}  is covariant under  $s$-independent gauge transformations and has fixed points at solutions to the classical Yang-Mills (Maxwell) equations of motion. When linearized for small fluctuations about a stable classical solution it behaves like the heat equation, damping out  the fluctuations  away from $s=0$.   An analogue  called Ricci flow was introduced by mathematicians over 50 years ago to smooth out metric fields while preserving diffeomorphism invariance  \cite{eells1964harmonic,hamilton1982three,perelman2002entropy} and was subsequently applied to gauge fields \cite{atiyah1983yang}.   The extension to quantum field theory employed here was developed  in Refs.~\cite{ Narayanan:2006rf,Luscher:2010iy,Luscher:2011bx}, with precursors in Refs.~\cite{Morningstar:2003gk,gusken1990study}, and has found a variety of useful applications in lattice QCD (see for example  \cite{Luscher:2013vga,DelDebbio:2013zaa, Suzuki:2013gza}).   

The effect of the flow equation \eqB{flow} is  illustrated by considering a $U(1)$ gauge field  in two Euclidean spacetime dimensions, flowing into a third dimension. We can decompose $ A_\mu$ as
\beq
 A_\mu = \partial_\mu  \omega +\epsilon_{\mu\nu}\partial_\nu \lambda ,
\eqn{decomp}\eeq
and extend $\omega$, $\lambda$ to $\bar\omega$, $\bar \lambda$ with boundary conditions $\bar\omega(p,s)\vert_{s=0} = \omega(p)$, $\bar\lambda(p,s)\vert_{s=0}  = \lambda(p)$ and solutions
\beq
\bar\omega(p,s) = \omega(p)\ ,\qquad \bar\lambda(p,s) = e^{-\xi p^2 |s|/\Lambda}\lambda(p)\ .
\eqn{u1sol}\eeq
Evidently the gauge degree of freedom $\bar\omega$ is constant over the entire extra dimension and can interact with the zeromodes at either domain wall if their individual gauge currents are not conserved, while the physical gauge field $\bar\lambda$ -- which is invariant under gauge transformations -- interacts  at full strength with the chiral zeromodes at $s=0$, but interacts with the mirror fermions at $s=\pm L$ with a Gaussian damping factor $\exp(-p^2/\mu^2)$, where
$
\mu \equiv  \sqrt{ \Lambda/\xi L}\ $ is an IR scale with $\mu/\Lambda\to 0$ as $L\to\infty$.   

It is helpful to regard this damping factor not as a property of the gauge field, but rather as a property of the fermions which behave as large objects  with a Gaussian form factor, incapable of reacting to gauge bosons attempting to transfer momentum $p\gg \mu$.
In fact, if there is an infrared cutoff on the $d$-dimensional space so that all gauge boson modes satisfy $p\gg\mu$, where $\mu$ can be made arbitrarily small, then  in the appropriate sequence of limits of vanishing $\mu$ and infinite volume, the mirror fermions can be made to decouple from the physical gauge field plane waves completely.  We will refer to the mirror fermions as ``fluff" because of their soft form factor.  \footnote{ In the lattice theory, exact symmetry of the continuum formulation that ensures massless chiral surface modes at finite $L$ is broken by the Wilson term, and the limit $\Lambda L \rightarrow \infty$ must also be taken in order to eliminate  residual $O(e^{-\Lambda L})$ chiral symmetry breaking couplings between fermions and fluff.}.

Even if we decouple the fluff, we still must ask whether the fermions in the bulk make contributions to the action which do not look $d$-dimensional. Naively it would seem that   the heavy bulk fermions and Pauli-Villars fields would decouple, only contributing local operators to the effective action suppressed by powers of their mass $\Lambda$, but that is incorrect; we know from  the analysis of  Callan and Harvey \cite{Callan:1984sa} that these modes generate a Chern-Simons term when integrated out, a marginal operator that  depends on the sign of $\Lambda$ but not its magnitude.  
This operator in the presence of an arbitrary $(d+1)$-dimensional background gauge field  $\bar A_\mu$ is, for $(d+1)=3,5$:   
\beq
S_3^\text{bulk}& =& c_3\frac{\Lambda}{|\Lambda|}\int \,  \left[\epsilon(s)-1\right] \Tr\left(  \bar F  \bar A - \textstyle{\frac{1}{3}} \bar A^3\right)\ ,\cr&&\cr
S_5^\text{bulk}& =& c_5 \frac{\Lambda}{|\Lambda|}\int \,\left[\epsilon(s)-1\right] \cr&&\qquad\times\Tr\left(\bar  F^2  \bar A
 -\half  \bar  F   \bar A ^3   +\textstyle{\frac{1}{10}}  \bar A^5\right).
\eqn{CS}
\eeq
where we are using $p$-form notation with $\bar A= \bar A_\mu^a T^a dx_\mu$, $ \bar F = d\bar A + \bar A^2$, $\mu=1,\ldots,d+1$, and 
\beq
c_{2n+1} = \frac{i^n}{2^{n+1} \pi^n (n+1)!}\ ,
\eeq
and our convention for $\gamma$ matrices is $\Tr \gamma_1\cdots\gamma_{2n+1} = (i)^n$.
  We now restrict these gauge fields $\bar A_\mu$  to the solution of \eq{flow}, with vanishing component in the $d+1$ direction, in which case only terms that  involve one derivative with respect to $s$ contribute to the above expression.  Note that as $(-\text{sgn } \Lambda)$ is the chirality of the zeromode at $s=0$ and 
$T^a$ are generators  in the same representation of the gauge group as the zeromode,  the sum of contributions to $S_{d+1}$ will cancel under the same algebraic condition as the vanishing of the  $d$-dimensional gauge anomaly among the zeromodes at $s=0$.   

The  variation of the above operators under gauge transformations  are total derivatives with respect to $s$, and integration over $s$  yields the consistent anomaly on the surfaces $s=0$ and $s=L$ after integration by parts, using the fact that $\partial_s\epsilon(s) = 2[\delta(s)-\delta(s-L)]$.  In particular, for a gauge transformation $ \Omega = \exp{i\varepsilon(x)}$, we find
\beq
\frac{\partial S^\text{bulk}_3}{\partial\varepsilon } &=& ic_3\frac{\Lambda}{|\Lambda|}\epsilon_{\mu\nu} \partial_\mu \bar A_\nu\Biggl\vert_{s=0}^{s=L} \cr
\frac{\partial S^\text{bulk}_5}{\partial\varepsilon } &=&  i c_5\frac{\Lambda}{|\Lambda|} \epsilon_{\mu\nu\rho\sigma}\,\partial_\mu\left[\bar A_\nu \bar A_\rho \bar A_\sigma  + 2 \bar A_\nu \partial_\rho  \bar A_\sigma\right]\Biggl\vert_{s=0}^{s=L} 
\eeq
which has exactly the right structure to cancel  the anomalies of the chiral modes at $s=0$ and at $s=\pm L$, each with the  correct local gauge field $\bar A_\mu(s)$,  which is precisely what is needed to account for the overall gauge invariance of $\Delta$. 
 
 The existence of the Chern-Simons operators \eq{CS} in the effective action precludes   interpreting the theory \eq{smeasure} as a local $d$-dimensional field theory.  To illustrate this we return to the simple $U(1)$ example in $d=2$ given in \eq{u1sol}.  In this case
 \beq
&&S_3^\text{bulk}\propto \int d^2x\int ds  \left[1-\epsilon(s)\right] \epsilon_{abc} \bar A_a\partial_b \bar A_c\cr
&&=2\int d^2x d^2y \left(\frac{\partial_\mu \partial_\alpha}{\Box} A_\alpha(x) \right)\Gamma(x-y)
 \left(\frac{\partial_\mu \partial_\beta}{\square}\epsilon_{\beta\gamma} A_\gamma(y)\right)  \cr &&
 \eeq
 with 
$\Gamma(r) = [\delta^2(r)-\frac{\mu^2}{4\pi}e^{-\mu^2 r^2/4} ]$, where we used the decomposition \eq{decomp} and solutions \eq{u1sol}. Because of the inverse Laplacian factors, the only way we can have the effective action behave like a local 2-dimensional theory is if either $\Gamma$ or the prefactor of $S_3$ vanishes. If we take the limit $\xi\to 0$ to turn off the gradient flow, then $\mu=\sqrt{\Lambda/\xi L}\to\infty$, $\frac{\mu^2}{4\pi}e^{-\mu^2 r^2/4} \to \delta^2(r)$,  and $\Gamma$ vanishes.  In this limit the gauge field has neither an $s$ component nor $s$ dependence and so cannot contribute to the Chern-Simons action.   This is the limit in which one recovers the conventional application of domain wall fermions to $d=2$ QED: one has a local, $2$-dimensional  theory of a massless Dirac fermion and vanishing Chern-Simons action. However, in the case we are interested in with  $\xi= 1$, then $\Gamma$ does not vanish in the limits $\mu=\sqrt{\Lambda/L}\to 0$   as $L\to\infty$,  and  the only way to recover an effective action with a local   2-dimensional description is to have the contributions of the various species of bulk fermions  to the Chern-Simons action  cancel, which is precisely equivalent to requiring the  fermion representation of the target chiral gauge theory in two dimensions to be free of gauge anomalies. With  the Chern-Simons operator vanishing, the remaining bulk fermion contributions to the effective action are suppressed by powers of $\Lambda$ and irrelevant.  This argument holds for the construction of chiral gauge theories in four dimensions as well.

  Up to this point we have only discussed a continuum model.  There are no apparent obstacles to discretizing the theory using an action similar to the one commonly used for domain wall fermions \cite{Shamir:1993zy}, only with gauge fields defined by \eq{flow}.  The extra dimension and the required large volume limits will pose computational challenges, but the biggest obstacle will likely be the sign problem that is generic for chiral gauge theories due to the phase of the integration measure in the continuum \cite{AlvarezGaume:1985di}.

\section{Topology}

Up to this point our analysis has focused on gauge field flow to the trivial fixed point of \eq{flow}, where $\bar A_\mu$ is pure gauge.  In general, any classical solution to the $d$-dimensional Yang-Mills (Maxwell) action is a fixed point, although it is plausible that the only attractive fixed  points   in each topological sector are the exact multi-instanton solutions.  For example, an arbitrary 4-dimensional Yang-Mills gauge field configuration $A_\mu(x,0)$ with winding number $k$   could be described by a field with fluctuations about  $n$ instantons and $\bar n = (n-k)$ anti-instantons, which  would   flow to a configuration at $s=\infty$ with  $k$ instantons and no anti-instantons. This would allow nontrivial correlation functions between ordinary fermions and fluff through 't Hooft interactions  \cite{'tHooft:1976up}.  For example, with one flavor of Dirac fermion $\psi$ and its fluff partner $\chi$, one would find a nonzero expectation value in such a gauge configuration for $(\mybar\psi_L\psi_R)^n(\mybar\psi_R\psi_L)^{\bar n} (\mybar\chi_L\chi_R)^{n-\bar n}$.  In a weakly coupled theory this operator would be proportional to $\exp[-(n+\bar n) S_0]$, where $S_0\propto1/\alpha$ is the large action for a single instanton, and the effect of the 't Hooft vertex would be negligible.  In a strongly coupled theory one would not expect topological effects to be suppressed; however it seems plausible that  a  generic configuration in volume $V$ with a large number of instantons and anti-instantons   $n,\bar n  \propto V$ but a much smaller net topological charge,  $(n-\bar n)\propto \sqrt{V}$, the spatial locations of the $(n-\bar n)$ instantons that survive the flow to $s=\infty$ would not be  highly correlated spatially with the $s=0$ gauge field configuration.  Thus the 't Hooft operator that received an expectation value would be of the form
$
\int d^4x\, \CO(x)\int d^4y \,\CO'(y),  
$
where $\CO$ is comprised of fermions and $\CO'$ of fluff. Such a vertex is nonlocal, but does not allow momentum to be transmitted between the two worlds at $s=0$ and $s=\infty$, and may not even be an extensive contribution to the action.  Thus  its phenomenology may prove to be benign.  A natural first step toward better understanding our proposal would be to investigate the nature of this topological interaction in a vectorlike gauge theory (where both matter and fluff are in a real representation of the gauge group) which would not suffer a sign problem. Other features of the theory could also be explored, such as whether colored fluff is confined.
  
If compatible with standard model phenomenology, the existence of fluff with its topological interactions could conceivably be a necessary, if unexpected, nonperturbative feature of chiral gauge theories. If one indeed takes such a view, then one must speculate whether the construction outlined in this Letter  is more than a prescription for the nonperturbative regularization of chiral gauge theories, or a technical approach  for their numerical simulation.  Could it  actually be realized in nature?   In this scenario the standard model possesses fluff  quarks and leptons which have resisted discovery due to their infinitely soft form factors under gauge  interactions (and possibly gravitational interactions, due to Ricci flow).   The prospect that fluff could decouple from propagating gauge fields yet participate in the topological structure of the vacuum is intriguing, perhaps allowing a massless fluff quark to solve the strong $CP$ problem without being easily seen.  The phenomenology and cosmology of such  matter is under investigation. 
     
 \begin{acknowledgments}
We gratefully acknowledge comments by M. Golterman, M. L\"uscher, R. Narayanan, and Y. Shamir.  This   work was supported by the NSF Graduate Research Fellowship under Grant No. DGE-1256082 and   by the DOE Grant No. DE-FG02-00ER41132.
\end{acknowledgments}

\bibliography{CGT.bib}
\end{document}